\documentclass[english,keywords,aps,amsmath,amssymb,twocolumn]{revtex4-1}
\usepackage[T1]{fontenc}
\usepackage[latin1]{inputenc}
\usepackage{babel}
\usepackage{amssymb}
\usepackage{graphicx}
\usepackage{color}
\usepackage{bm}
\usepackage{longtable}
\usepackage{amsmath} 
\usepackage{amsfonts}
\usepackage{amssymb}
\begin{document}
\title{Joint weak value for all order coupling using continuous variable and qubit probe}
\author{Asmita Kumari$^{1}$}
\author{A. K. Pan$^{1}$ \footnote{akp@nitp.ac.in}}
\author{P. K. Panigrahi$^{2}$}
\affiliation{$^{1}$ National Institute Technology Patna, Ashok Rajhpath, Patna, Bihar 800005, India}
\affiliation{$^{2}$Indian Institute of Science Education and Research Kolkata, Mohanpur, Nadia 741246, India}

\begin{abstract}
The notion of weak measurement in quantum mechanics has gained a significant and wide interest in realizing apparently counterintuitive quantum effects. In recent times, several theoretical and experimental works have been reported for demonstrating  the joint weak value of two observables where the coupling strength is restricted to the second order. In this paper, we extend such a formulation by providing a complete treatment of joint weak measurement scenario for all-order-coupling for the observable satisfying $A^2=\mathbb{I}$ and $A^2=A$, which allows us to reveal several hitherto unexplored features. By considering the probe state to be discrete as well as continuous variable, we demonstrate how the joint weak value can be inferred for any given strength of the coupling. A particularly interesting result we pointed out that even if the initial pointer state is uncorrelated, the single pointer displacement can provide the information about the joint weak value, if at least third order of the coupling is taken into account. As an application of our scheme, we provide an all-order-coupling treatment of the well-known Hardy paradox by considering the continuous as well as discrete meter states and show how the negative joint weak probabilities emerge in the quantum paradoxes at the weak coupling limit.

\end{abstract}
\maketitle
\section{Introduction}
The path-breaking idea of weak measurement (WM) in quantum mechanics(QM), originally proposed by Aharonov, Albert and Vaidman (AAV) \cite{aav}, has gained wide interest in realizing apparently counterintuitive quantum effects. Contrary to the strong measurement, in WM scenario, the average value of an observable (coined as weak value) can yield results beyond the eigenvalue spectrum of the measured observable. In last decade, a flurry of works have been reported, in which the WM and its implications have been extensively studied, both theoretically \cite{duck,av91,mit,jeff,jozsa,brunner,vaidman,aha02,resch,lundeenhardy,lorenzo,panmatz,nakamura,nisi,puentes,shikano,bernardo,kanjilal,vaid13} and experimentally \cite{ritchie,pryde,wang,hosten,jwvexp,starling,steinberg11,lundeen11,zil,danan,kim,str,piacentini}.  In one hand, WM provides new insights into conceptual quantum paradoxes \cite{av91,vaidman,aha02,mit,vaid13,danan} and on the other hand, it provides several practical applications, such as, identifying tiny spin Hall effect \cite{hosten}, detecting very small transverse beam deflections \cite{starling}, measuring average quantum trajectories for photons \cite{steinberg11}, improving signal-to-noise ratio for determination of small phase through interferometry \cite{str} and protecting a quantum state \cite{kim}. 

Before proceeding further, let us first encapsulate the notion of WM in QM. The WM scenario comprises three steps; preparation of the system in a quantum state (commonly termed as pre-selection), an intermediate perturbation for producing the weak correlation between the system and pointer states, and the selection of a suitable sub-ensemble corresponding to a particular quantum state (commonly known as post-selection). Let the $|\chi_i\rangle$ be the pre-selected state prepared by a suitable strong measurement, and  $|\psi\rangle=\int \psi(x)|x\rangle dx$ be the initial pointer state in position space. Next step is to introduce a perturbation, $H_{I}= g(t) \hat A\otimes \hat P_{x}$, where $g(t)$ is a smooth function of $t$ obeying $\int_{0}^{t}g(t)dt= g$, and $t$ is the time during which the interaction takes place. The system-pointer total state after the perturbation can then be written as $|\Psi^{\prime}\rangle=\int   e^{-ig \hat A\otimes \hat P_{x}}\psi(x)|x\rangle|\chi_{i}\rangle dx$. The final and the crucial step is to invoke a suitable strong measurement to post-select a sub-ensemble corresponding to a system state $|\chi_{f}\rangle$. Due to the post-selection, the pointer is then left in a state, $|\psi_{f}\rangle =\langle\chi_{f}|\Psi^{\prime}\rangle=\int  \langle \chi_{f}| e^{-ig  \hat A\otimes\hat P_{x}}\psi(x)|x\rangle|\chi_{i}\rangle dx$. Now, if $g$ is sufficiently small (i.e., the perturbation is taken to be weak), one can neglect the higher order terms involving $g$. In such a case, the final pointer state can be approximated as $|\psi_{f}\rangle = \langle \chi_{f}|\chi_{i}\rangle  \int   e^{-ig \hat{A_{w}}\otimes\hat P_{x}}\psi(x)|x\rangle dx$, where, 
\begin{equation}
\hat{A}_{w}=\frac{\langle \chi_{f}| \hat A|\chi_{i}\rangle}{\langle \chi_{f}|\chi_{i}\rangle}
\end{equation} 
is defined as the weak value of the observable $\hat A$ and $|\langle \chi_{f}|\chi_{i}\rangle|^{2}$ is the post-selection probability. For  suitable choices of $|\chi_{f}\rangle$ and $|\chi_{i}\rangle$, $\hat{A}_{w}$ can be made far outside the eigenvalue ranges of the observable $\hat{A}$, at the expanse of the small post-selection probability. Interestingly, the weak value can also be complex \cite{jozsa}, yet not unphysical. For example, for Gaussian pointer, the imaginary part of the complex weak value  produces the pointer shift in the space in which the perturbation is introduced, and the real part produces the pointer shift in the conjugate space. For the above mentioned interaction Hamiltonian, $\langle \hat X\rangle_{f}-\langle \hat X\rangle_{i} \propto \Re(\hat{A}_{w})$ and  $\langle \hat P_x\rangle_{f}-\langle \hat P_x\rangle_{i} \propto \Im(\hat{A}_{w})$, where $\langle \hat X\rangle_{f}$ denotes the average value of the pointer observable in the post-selected state, and similarly for others. 

Note that, this simple correspondence between the pointer shifts and weak values breaks down, if the higher order expansion of the perturbation is considered, instead of the commonly used first order approximation. This is due to the fact that the retention of the higher order expansion terms of the perturbation may produce many peaks \cite{panmatz} in the probability distribution of the post-selected meter state.  Thus, for higher order expansion, the expression appears in the form of $\langle \chi_{f}| \hat A|\chi_{i}\rangle/\langle \chi_{f}|\chi_{i}\rangle$ of an observable $\hat A$ is merely a conditional average value which has no bearing on the weak value in the usual sense of WM. For historical good reason, we  will be following the conventional terminology. The exact treatment of the AAV setup for all-order-coupling has been discussed in Ref. \cite{lorenzo,nakamura,panmatz} and some interesting results like semi-weak and orthogonal weak values have been pointed out. 

The purpose of this paper is to study the joint weak value for all-order-coupling. In recent times, the joint weak value of two observables has been studied \cite{resch,vaid,wuz,puentes,shikano,bernardo,kanjilal} and experimentally verified   \cite{jwvexp,piacentini}. Among other intriguing implications, the measurement of joint weak value is particularly interesting because a strong von Neumann measurement of joint mean values requires a nonlinear Hamiltonian, which is a difficult task to achieve \cite{truchette}. Resch and Steinberg \cite{resch} first introduced  a scheme, where  the joint weak value of two commuting observables (say, $\hat{A}$ and $\hat{B}$), can be obtained from the statistics of the correlations between pointer variables. Specifically, they found that the joint expectation value of the pointer observables is proportional to the $(\hat{A}\hat{B})_{w}$, if second order expansion of the coupling is taken into account.  

We note here that in all the existing literature demonstrating joint weak value, the coupling strength is restricted to the second order. In this paper, we extend the earlier formulations (mainly, the work of Resch and Steinberg \cite{resch}) by providing a complete treatment of joint weak measurement scenario for all-order-coupling, which allows us to reveal several unexplored features. In order to doing this, we consider two types of observables satisfying $A^2=\mathbb{I}$ and $A^2=A$ by considering both continuous and discrete pointer states. We demonstrate that the joint weak value can be extracted for an arbitrary strength of the coupling and the results of Resch and Steinberg \cite{resch} can be recovered if expansion of the coupling is taken to be up to the second order. We have also demonstrated the single pointer displacement can provide the information about the joint weak value if at least third order of the coupling is taken into account. This is particular interesting because the initial pointer we consider is uncorrelated. It is also shown that real and imaginary part of single and joint weak value can be extracted from the statistics of single pointer observable. As an application of our treatment, we reexamine the well-known Hardy paradox \cite{hardy} and show how negative probabilities emerges at the weak measurement regime.

 The paper is organized as followed. In Section $\mathbb{II}$, we demonstrate how joint weak value can be extracted for projector and general observables for all order of coupling by taking the continuous pointer and in Section $\mathbb{III}$, we calculate the same by using discrete pointer state. We then consider the all order coupling treatment to examine the Hardy's paradox in Section $\mathbb{IV}$. We give a brief summary of our findings in Section $\mathbb{V}$.

\section{Joint Weak Value using continuous variable probe state}
Let us first explore the joint weak measurement scenario by using the continuous variable pointer state. For this, we consider the system-apparatus state at $t=0$, is given by,
\begin{equation}
|\Psi_{i}\rangle=|\psi_{i}\rangle|\chi_{i}\rangle=\int \psi(x,y)|x,y\rangle |\chi_{i}\rangle dx dy,
\end{equation}
where $|\chi_{i}\rangle$ is the pre-selected system state and $\psi(x,y)$ is the uncorrelated pointer wave function at $t=0$, which is taken to be a two-dimensional Gaussian is of the form
\begin{eqnarray}
\label{gauss}
\psi(x,y)=\left(2 \pi \sigma^{2}\right)^{-1/2}exp\left(-\frac{x^2 +y^2}{4\sigma^2}\right),
\end{eqnarray}
where $\sigma$ is the half-width of the Gaussian wave packet.
 Let us now introduce the perturbation is of the form $\hat{H}=g(\hat A \otimes\hat P_x + \hat B \otimes\hat P_y)$, where $\hat A$ and $\hat B$ are observables of the system such that, $[\hat A,\hat B]=0$,  $\hat P_{x}$ and $\hat P_y$ are the conjugate momenta of the pointer position observables $\hat X$ and $\hat Y$ respectively. The term $g=\int_{0}^{t} g(t) dt$ determines the weakness of the perturbation.  In this paper, we consider two different types of system observables satisfying $A^2=\mathbb{I}$ and $A^2=A$. Now, the post-interaction system-apparatus state can be written as  
\begin{eqnarray}
|\Psi\rangle=\int  \  e^{-i g(\hat A \otimes\hat P_x + \hat B \otimes\hat P_y)}\psi(x,y)|x,y\rangle|\chi_{i}\rangle dx dy,
\end{eqnarray}

 By invoking the post-selection in the state $|\chi_f\rangle$, the pointer we have

\begin{eqnarray}
\nonumber
|\psi_{f}\rangle=\int \langle\chi_{f}| \  e^{-i g(\hat A \otimes\hat P_x + \hat B \otimes\hat P_y)}\psi(x,y)|x,y\rangle|\chi_{i}\rangle dx dy .\\
\end{eqnarray}

The post-selected pointer state will be used to evaluate the pointer displacements of various pointer observables.

\subsection{Joint weak value of the observables $\hat A^2=\mathbb{I}$ and $\hat B^2=\mathbb{I}$}
In order to obtain the joint weak value of the observables $\hat{A}$ and $\hat{B}$ satisfying $\hat A^2=\mathbb{I}$, $\hat B^2=\mathbb{I}$ and  $[\hat A,\hat B]=0$ we expand the interaction Hamiltonian for all-order-coupling. The post-selected pointer state can then be written as,
\begin{eqnarray}
\label{pointer}
|\psi_{f}\rangle=\langle\chi_{f}|\chi_{i}\rangle\int  \gamma_{1} \psi(x,y)|x,y\rangle dx dy ,
\end{eqnarray}

where
 $\gamma_{1}=\cos (\hat P_x g) \cos (\hat P_y g)-i(\hat{A})_{w}\sin (\hat P_x g) \cos (\hat P_y g)-i (\hat{B})_{w} \sin (\hat P_y g) \cos (\hat P_x g) -(\hat{A}\hat{B})_{w} \sin (\hat P_x g) \sin (\hat P_y g)$. Here, $(\hat{A})_{w}$ and $(\hat{B})_{w}$ are weak values of the system observables $\hat{A}$ and $\hat{B}$ respectively and $(\hat{A}\hat{B})_{w}$ represents the joint weak value. \\

 Next, in order to calculate average pointer displacement, we consider a pointer observable say, $\hat M$. The mean value of  $\hat M$ on the post-selected state $|\psi_{f}\rangle$ is defined as,
\begin{eqnarray}
\langle \hat M\rangle_{f}=Tr\left[\hat M \frac{|\psi_{f}\rangle\langle\psi_{f}|}{\langle\psi_{f}|\psi_{f}\rangle}\right],
\end{eqnarray}
where $\hat M$ can be single or joint meter observables. The explicit form of $\langle \hat M\rangle_{f}$ for Eq.($\ref{pointer}$) is given by

\begin{eqnarray}
\label{M}
\langle \hat M\rangle_{f}&=&\Big[\langle  H_1^{\dagger} \hat M H_1\rangle_{i}-i\langle  H_1^{\dagger} \hat M H_2\rangle_{i}\\
\nonumber
&+&i\langle  H_2^{\dagger} \hat M H_1\rangle_{i}+\langle  H_2^{\dagger} \hat M H_2\rangle_{i}\Big]W^{-1} ,
\end{eqnarray}
where $\langle..\rangle_{i}=\langle \psi_{i}|..|\psi_{i}\rangle$,   
$H_1=\cos  (\hat P_x g) \cos  (\hat P_y g)- (\hat{A}\hat{B})_{w}\sin (\hat P_x g) \sin ( \hat P_y g)$, $H_2= (\hat{B})_{w}\cos ( \hat P_x g) \sin ( \hat P_y g)+ (\hat{A})_{w}\sin ( \hat P_x g) \cos ( \hat P_y g)$ and $W={\langle\psi_{f}|\psi_{f}\rangle}/Pr =[\langle  H_1^{\dagger} H_1\rangle-i\langle  H_1^{\dagger} H_2\rangle+i\langle  H_2^{\dagger}  H_1\rangle+\langle  H_2^{\dagger}  H_2\rangle]$ with the post-selection probability($Pr$) is given by $Pr=|\langle\chi_{f}|\chi_{i}\rangle|^2$.

Note that, in writing Eq.(\ref{M}), we have made \emph{no} approximation on the coupling strength. By using Eq.(\ref{gauss}), the joint mean value of two point observables (say, $\hat M= \hat{X}\hat{Y}$) can be calculated. The joint pointer displacement is then given by
\begin{eqnarray}
\label{xy}
\langle \hat X \hat Y\rangle_{fi}= g^2  \left(\Re[(\hat{A}\hat{B})_{w}]+\Re[(\hat{A})_{w}^{\ast}(\hat{B})_{w}]\right) W^{-1}_{1}, 
\end{eqnarray}
where  
\begin{eqnarray}
\nonumber
W_1=  e^{-\frac{g^2}{\sigma ^2}}\left[c_1 + |(\hat{A}\hat{B})_w|^2 c_2-\left(|(\hat{B})_w|^2 +|(\hat{A})_{w}|^2\right)c_3\right]
\end{eqnarray}

with $c_1=(1 + e^{\frac{g^2 }{2\sigma^2}})^2$,  $c_2=(1 - e^{\frac{g^2 }{2\sigma^2}})^2$ and $c_3=(1 - e^{\frac{g^2 }{\sigma^2}})$.

Here the $\langle ..\rangle_{fi}= \langle ..\rangle_{f}-\langle ..\rangle_{i}$ mean displacement of pointer observable. Note that, Eq.(\ref{xy}) contains the term $\Re[(\hat{A}\hat{B})_{w}]$, thereby implying that we can access the joint weak value for any strength of the coupling. Interestingly, Eq.(\ref{xy}) in its compact form, appears similar to the expression of the paper by Resch and Steinberg \cite{resch} but differed by the term $W_1^{-1}$. It can easily be seen from Eq.(\ref{xy}) that for first order expansion, $\langle \hat X \hat Y\rangle_{fi}=0$. However, for the second order expansion, we can exactly recover the joint weak value expression of Resch and Steinberg \cite{resch},
\begin{equation}
\label{rs}
\Re[(\hat{A}\hat{B})_{w}]= \frac{2}{g^2} \langle \hat X  \hat Y\rangle_{fi}-\Re[(\hat{A})_{w}^{\ast}(\hat{B})_{w}] ,
\end{equation}
As mentioned in \cite{resch}, in order to extract joint weak value $\Re[(\hat{A}\hat{B})_{w}]$ from Eq.(\ref{rs}), the weak values of both $(\hat{A})_w$ and $(\hat{B})_w$ need to obtained from separate experiments.

 In order to find imaginary part of joint weak value for all-order-coupling we choose joint mean value of two meter observables, $\hat M= \hat{X} \hat P_{y}$, so that,
\begin{eqnarray}
\label{xpc}
\langle \hat{X} \hat P_{y}   \rangle_{fi}=e^{-\frac{g^2}{2 \sigma ^2}}\frac{g^2  (\Im[(\hat{A})^*_{w}(\hat{B})_{w}] +\Im[(\hat{A}\hat{B})_w])}{2 \sigma ^2 W{_2} },
\end{eqnarray} 
with
\begin{eqnarray}
W_2&=&\Big[|(\hat{A}\hat{B})_w|^2-(|(\hat{A}\hat{B})_w|^2+1) \cosh \left(\frac{g^2}{2 \sigma ^2}\right)\\
\nonumber
&-&\sinh \left(\frac{g^2}{2 \sigma ^2}\right) (|(\hat{A})_{w}|^2+|(\hat{B})_{w}|^2)-1 \Big] e^{\frac{g^2}{2 \sigma ^2}} .
\end{eqnarray}

Then the imaginary part of joint weak value $\Im[(\hat{A}\hat{B})_w]$ can also be obtained from our all-order-coupling. At the second order coupling limit, Eq.($\ref{xpc}$) will look similar to Eq.($\ref{xy}$).\\

 Next, we show an important result that the joint weak value can be extracted from the statistics of a single pointer observable. It is particularly interesting because our initial pointer state is uncorrelated, but all order coupling treatment produces a kind of correlation which enables us to obtain joint weak value from single pointer statistics. For showing this, we calculate the mean displacement of the pointer observable, $\hat M=\hat X$, is of the form,
\begin{eqnarray}
\label{meanx}
\langle \hat X  \rangle_{fi}&=& 2 g (( \Re[(\hat{A})_w]- \Re[(\hat{B})_{w} (\hat{A}\hat{B})_w^{\ast}])e^{-\frac{g^2 }{2 \sigma^2}} \\
\nonumber
&+&( \Re[(\hat{A})_w] + \Re[(\hat{B})_{w} (\hat{A}\hat{B})_w^{\ast}])  )W^{-1}_{1},
\end{eqnarray}
where $\langle \hat X \rangle_{i}=0$. It is thus evident from Eq.(\ref{meanx}) that, the joint weak value can indeed be inferred from the displacement $\langle \hat X\rangle_{fi}$, (instead of $\langle \hat X\hat Y\rangle_{fi}$ used in \cite{resch,shikano}). Note that, for the first and second order expansions of the coupling, we have $\langle \hat X \rangle_{fi}= g\Re[(\hat{A})_w]$, giving information about the observable $\hat A$ only. But if the coupling strength is taken upto third order, we then have
\begin{eqnarray}
\label{meanx2}
\langle \hat{X}  \rangle_{fi}&=& g\Re[(\hat{A} )_w]  +\frac{g^3}{4 \sigma^2} \Big(\Re[(\hat{A})_{w}](1-|(\hat{A})_{w}|^2\\
\nonumber
&-&|(\hat{B})_{w}|^2)+\Re[(\hat{B})^*_{w}(\hat{A}\hat{B})_w]\Big),
\end{eqnarray}
Hence, the single pointer displacement can provide the information about joint weak value, if at least third order coupling is taken into account.
In fact the real and imaginary part of $(\hat{B})_{w}$ can also be obtained from the pointer even if the observable $\hat{B}$ is not initially associated with the pointer variable $\hat{X}$. For simplicity, if one takes $(\hat{A} \hat{B})_w$, $(\hat{A})_{w}$ and $(\hat{B})_{w}$ as all real, then using Eq.(\ref{meanx2}), the joint weak value $(\hat{A}\hat{B})_w$ can be written as

\begin{eqnarray}
\label{ab}
\Re[(\hat{A}\hat{B})_w] &=&\frac{4 \sigma^2 \langle \hat X \rangle_{fi}}{g^3 (\hat{B})_{w}}- \frac{4 \sigma^2}{g^2} \frac{(\hat{A})_{w}}{(\hat{B})_{w}}\\
\nonumber
& &-\frac{(\hat{A})_{w}(1-|(\hat{A})_{w}|^2-|(\hat{B})_{w}|^2)}{ (\hat{B})_{w}} ,
\end{eqnarray} 
As is required in Resch and Steinberg \cite{resch}, in order to obtain $(\hat{A}\hat{B})_w$ in our case, the values of $(\hat{A})_{w}$ and $(\hat{B})_{w}$ need to be obtained from other measurements. This feature was not earlier noticed due to the lack of all-order-coupling treatment of the joint weak measurement.

 A simple example can be useful to understand the usefulness of the result. For this we take initial system state as $|\psi_{i}\rangle=|{+_z}\rangle\otimes|{+_z}\rangle$ and post-selected system state as $|\psi_{f}\rangle=(\cos(\theta)|{+_z}\rangle+i\sin(\theta)|{-_z}\rangle)\otimes|{+_z}\rangle $ where $|{+_z}\rangle$ is eigenstate of pauli observable $\sigma_z$. The system observables are $\hat{A}=\hat{\sigma}_{x}\otimes\mathbb{I}$ and $\hat{B}=\mathbb{I}\otimes\hat{\sigma}_{z}$. Then the single weak values $(\hat{A})_{w}=i \tan(\theta)$ and $(\hat{B})_{w}=1$. Substituting the weak values in Eq.(\ref{ab}), we found that joint weak value is given by
\begin{eqnarray}
(\hat{\sigma_{x}}\otimes\hat{\sigma_{z}})_w =\frac{4 \sigma^2 \langle \hat X \rangle_{fi}}{g^3 },
\end{eqnarray} 

Thus, for particular choice of system state and observables the joint weak value can simply be made proportional to mean displacement of single pointer observable if higher order coupling is considered. 
 
Next, in order to find the real and imaginary parts of joint mean value, we consider the mean value of ${\hat{X}}^2$. Then the pointer displacement can be written as
\begin{eqnarray}
\label{meanx3}
&&\langle {\hat{X}}^2 \rangle_{fi} = e^{\frac{g^2}{\sigma ^2}} \left(g^2+\sigma ^2\right) \Big(|(\hat{A})_{w}|^2+|(\hat{B})_{w}|^2\\
\nonumber
&+&|(\hat{A}\hat{B})_{w}|^2+1\Big)+e^{\frac{g^2}{2 \sigma ^2}}\Big(g^2 (|(\hat{A})_{w}|^2-|(\hat{B})_{w}|^2\\
\nonumber
&-&|(\hat{A} \hat{B})_{w}|^2+1)-2 \sigma ^2 (|(\hat{A} \hat{B})_{w}|^2-1)\Big)\\
\nonumber
&+&\sigma ^2 \Big(-|(\hat{A})_{w}|^2-|(\hat{B})_{w}|^2+|(\hat{A}\hat{B})_{w}|^2+1\Big),
\end{eqnarray}
 It can be seen from Eq.$(\ref{meanx3})$ that one can access both the imaginary and real weak values of the observables $\hat A$ and $\hat B$ from the single pointer displacement. For second order expansion, we have
\begin{eqnarray}
\langle {\hat{X}}^2 \rangle_{fi}&=&\sigma ^2 +\frac{1}{2} g^2 (|(\hat{A})_{w}|^2+1),
\end{eqnarray}
which contains only the $A_w$ but we can still access the real and imaginary part of single weak value. By taking fourth order expansion of the coupling, 
\begin{eqnarray}
&&\langle {\hat{X}}^2 \rangle_{fi}=\sigma ^2 +\frac{1}{2} g^2 (|(\hat{A})_{w}|^2+1)\\
\nonumber
&+& \Big(1- |(\hat{A})_{w}|^2(1-|(\hat{B})_{w}|^2)+|(\hat{A} \hat{B})_{w}|^2\Big)\frac{g^4 }{8 \sigma^2 }.
\end{eqnarray}

Then, real and imaginary part of joint weak value is accessible  from $\langle {\hat{X}}^2 \rangle_{fi}$. Note that, the mean displacement of  $\hat{X}^2$ also provides the information of $|(\hat{B})_{w}|^{2}$, as opposed to Ref. \cite{puentes} in which the authors claimed that the above feature cannot be obtained if initial pointer is taken to be Gaussian.

\subsection{Joint weak value for two projectors}
We now calculate the joint weak measurement for all-order-coupling for projectors. Following the similar procedure used in earlier we introduce the Hamiltonian  $\hat{H}=g(\hat{P_A}\otimes\mathbb{I} \otimes \hat P_x + \mathbb{I}\otimes\hat{P_ B} \otimes\hat P_y)$, where, $\hat{P_A}\otimes\mathbb{I}$ and $\mathbb{I}\otimes\hat{P_ B} $ are projectors. Using all-order-coupling treatment also in this case, the post-selected pointer state is given by
   \begin{eqnarray}
|\psi_{f}\rangle=\langle\chi_{f}|\chi_{i}\rangle\int \gamma_{2} \psi(x,y)|x,y\rangle dx dy  ,
\end{eqnarray}
where $\gamma_{2} =1-(\hat{P_A}\otimes\mathbb{I})_{w}(1-e^{-i g \hat P_{x}})-(\mathbb{I}\otimes\hat{P_B})_{w}(1-e^{-i g\hat P_{y}})+(\hat{P_A}\otimes\hat{P_B})_{w}(1-e^{-i g\hat P_{x}})(1-e^{-i g\hat P_{y}})$ in which $(\hat{P_A}\otimes\mathbb{I})_{w}$ and $(\mathbb{I}\otimes\hat{P_B})_{w}$ are weak value of projectors $\hat{P_A}\otimes\mathbb{I}$ and $ \mathbb{I}\otimes\hat{P_B}$ respectively and $(\hat{P_A}\otimes\hat{P_B})_{w}$ is their joint weak value.

The mean displacement of the joint pointer observable, $\hat M = \hat X \hat Y$, is given by
\begin{eqnarray}
\label{xyp}
\langle \hat{X} \hat{Y} \rangle_{fi} &=& \frac{g^2}{2 W_{3}} \Big[\Re[(\hat{P_A}\otimes\mathbb{I})^{*}_{w}( \mathbb{I}\otimes\hat{P_B})_{w}]+\Re[(\hat{P_A}\otimes\hat{P_B})_{w} ]\\
\nonumber
&+&2|(\hat{P_A} \otimes\hat{P_B})_{w}|^2e^{-\frac{g^2}{4 \sigma ^2}}-2\Re[(\hat{P_A}\otimes\mathbb{I})^{*}_{w}( \hat{P_A}\otimes\hat{P_B})_{w}] \\
\nonumber
&+&\Re[( \mathbb{I}\otimes\hat{P_B})_{w}(\hat{P_A} \otimes\hat{P_B})_{w}] -2|(\hat{P_A} \otimes\hat{P_B})_{w}|^2\Big]\\
\nonumber
&\times&(1-e^{\frac{g^2}{8 \sigma ^2}})) ,
\end{eqnarray}
which contains joint weak value term $(\hat{P_A}\otimes\hat{P_B})_{w}$.

Again for second order expansion of $\langle \hat{X} \hat{Y} \rangle_{fi}$ in Eq.($\ref{xyp}$),  we can recover the result of Resch and Steinberg \cite{resch} for joint pointer displacement 
\begin{eqnarray}
\nonumber
\langle \hat{X} \hat{Y} \rangle_{fi}=\frac{g^2}{2}(  \Re[(\hat{P_A}\otimes\mathbb{I})_{w}( \mathbb{I}\otimes\hat{P_B})^*_{w}]+\Re[(\hat{P_A}\otimes\hat{P_B})_{w}]),\\
\end{eqnarray}

The mean displacement of single pointer observable, $\hat M=\hat X$, can be obtained as
\begin{eqnarray}
\label{xp}
&&\langle \hat X  \rangle_{fi}= \frac{e^{-\frac{g^2 }{4 \sigma^2}}}{W_{3}}\Re[((\hat{P_A}\otimes\mathbb{I})^*_{w}( \mathbb{I}\otimes\hat{P_B})_{w}\\
\nonumber
&+&(\hat{P_A}\otimes\hat{P_B})_{w}-2( \mathbb{I}\otimes\hat{P_B})^*_{w}(\hat{P_A}\otimes\hat{P_B})_{w})(1-e^{\frac{g^2 }{8 \sigma^2}})\\
\nonumber
&+& 2(|(\hat{P_A}\otimes\hat{P_B})_{w}|^2 -(\hat{P_A}\otimes\mathbb{I})_{w}(\hat{P_A}\otimes\hat{P_B})_{w})\\
\nonumber
&\times&(1-2 e^{\frac{g^2 }{8 \sigma^2}})+(|(\hat{P_A}\otimes\mathbb{I})^*_{w}|^2-2(\hat{P_A}\otimes\mathbb{I})^*_{w}\\
\nonumber
&\times&(\hat{P_A}\otimes\hat{P_B})_{w}+2|(\hat{P_A}\otimes\hat{P_B})_{w}|^2)e^{\frac{g^2 }{4 \sigma^2}} ],
\end{eqnarray} 

 It can be seen from Eq.($\ref{xp}$) that all-order-coupling treatment enables us to provide joint weak value from the single observable for the case of projector too. For first and second order of expansion we obtain the well known result $\langle \hat X  \rangle_{f}= g\Re[(\hat{P_A}\otimes\mathbb{I})_w]$. But for third order of coupling, we have  
	
\begin{eqnarray}
\langle \hat X  \rangle_{fi}&=& g\Re[(\hat{P_A}\otimes\mathbb{I})_w]+\frac{g^3}{8 \sigma^2} \Re[(\hat{P_A}\otimes\mathbb{I})_{w}\\
\nonumber
&-&((\hat{P_A}\otimes\mathbb{I})_{w})^2-2|(\hat{P_A}\otimes\mathbb{I})_{w}|^2+2((\hat{P_A}\otimes\mathbb{I})_{w})^2\\
\nonumber
&\times&(\hat{P_A}\otimes\mathbb{I})^*_{w}-(\hat{P_A}\otimes\mathbb{I})_{w}( \mathbb{I}\otimes\hat{P_B})_{w}\\
\nonumber
&+&2(\hat{P_A}\otimes\mathbb{I})^*_{w}|( \mathbb{I}\otimes\hat{P_B})_{w}|^2+(\hat{P_A}\otimes\hat{P_B})_{w}\\
\nonumber
&-&2( \mathbb{I}\otimes\hat{P_B})^*_{w}(\hat{P_A}\otimes\hat{P_B})_{w}],
\end{eqnarray} 
Hence, if the coupling strength is taken upto the third order, joint weak value can also be obtained also for the projectors.\\

 Following the same procedure we obtain joint expectation value  $\langle\hat P_{x} \hat P_{y}\rangle_{fi}$ of pointer momentum given by
\begin{eqnarray}
\label{p1p2}
\nonumber
\langle \hat{ P}_{x} \hat{ P}_{y} \rangle_{fi}=\frac{g^2\Big( \Re [(\hat{P_A} \otimes\mathbb{I})_{w}( \mathbb{I}\otimes\hat{P_B})^{*}_{w}]-\Re  [(\hat{P_A}\otimes\hat{P_B})_{w}]\Big) }{4 \sigma ^{4} W_3 }.
\end{eqnarray}
where $W_{3}$ is normalization constant given in Eq.$(\ref{A1})$. The presence of joint weak value term from  Eq.$(\ref{xyp})$ and Eq.$(\ref{p1p2})$ imply that we can find joint weak value for all-order-coupling for projectors too,  which will be used in \textit{Sec}.$IV$ where we consider Hardy set-up \cite{hardy}.

\section{Joint Weak Value for qubit pointer for projectors}
 The motivation of studying joint weak measurement for the qubit pointer is the following. There have been many works \cite{vaidman, aha02} that use path and polarization (or spin) as a pointer or system state. Then the calculation of joint weak value using discrete meter can be useful to verify this issue experimentally.
 In order to calculate the joint weak value for qubit pointer, we consider the system-pointer state at $t=0$ as,
\begin{equation}
|\Xi_{i}\rangle=|\xi_{i}\rangle \otimes |\chi_{i}\rangle,
\end{equation}

where $|\xi_{i}\rangle$ is the pointer qubit pointer state and $|\chi_{i}\rangle$ is the pre-selected system state. The interaction Hamiltonian is taken to be, $\mathcal{H}_{I}=g(t)[( \hat{P_A}\otimes\mathbb{I})\otimes(\hat\sigma_{1}\otimes\mathbb{I}) +  (\mathbb{I}\otimes\hat{P_B})\otimes (\mathbb{I}\otimes\hat\sigma_{2})]$, where $\hat\sigma_{1}$ and $\hat\sigma_{2}$ are arbitrary satisfying ${\hat\sigma}^2_{1}=\mathbb{I}$ and ${\hat\sigma}^2_{2}=\mathbb{I}$. 
The post-interaction state can be written as,
\begin{eqnarray}
|\Xi_{f}\rangle =  e^{-ig [ (\hat{P_A}\otimes\mathbb{I})\otimes(\hat\sigma_{1}\otimes\mathbb{I}) + ( \mathbb{I}\otimes\hat{P_B})\otimes (\mathbb{I}\otimes\hat\sigma_{2})]}|\xi_{i}\rangle\otimes|\chi_{i}\rangle,
\end{eqnarray}
Using the property of projectors $(\hat{P_A}\otimes\mathbb{I})^2 = \hat{P_A}\otimes\mathbb{I}$, $( \mathbb{I}\otimes\hat{P_B})^2 =  \mathbb{I}\otimes\hat{P_B}$ and by considering the post-selected system state $|\chi_{f}\rangle$, the post-interaction pointer state can be written as
\begin{eqnarray}
|\xi_{f}\rangle=\langle\chi_{f}|\chi_{i}\rangle \eta_{1} |\xi_{i}\rangle,,
\end{eqnarray}
where,  $\eta_{1} =1-(\hat{P_A}\otimes\mathbb{I})_{w} (1-  \cos \left(g\right)+i (\hat\sigma_{1}\otimes\mathbb{I}) \sin \left(g\right))-( \mathbb{I}\otimes\hat{P_B})_{w}(1-\cos \left(g\right)+i (\mathbb{I}\otimes\hat\sigma_{2}) \sin \left(g\right))+ (\hat{P_A}\otimes\hat{P_B})_w(1-  \cos \left(g\right)+i (\hat\sigma_{1}\otimes\mathbb{I}) \sin \left(g\right))(1-\cos \left(g\right)+i (\mathbb{I}\otimes\hat\sigma_{2}) \sin \left(g\right))$. 
The mean value of a suitable pointer observable, say, $(\hat M =\hat\sigma_{1} \otimes \hat\sigma_{2} )$, on the post-selected state $|\xi_{f}\rangle$ is given by,

\begin{widetext}
\begin{eqnarray}
\label{Mq}
\langle \hat M \rangle_{fi} &=& \Big[2\Re[(\hat{P_A}\otimes\mathbb{I})_w( \mathbb{I}\otimes\hat{P_B})^{*}_w + (\hat{P_A}\otimes\hat{P_B})_w]\sin ^2(g)-\Im[(\hat{P_A}\otimes\mathbb{I})^{*}_w(\hat{P_A}\otimes\hat{P_B})_w+2(\mathbb{I}\otimes\hat{P_B})_w \\
\nonumber
&+& 2 r (\hat{P_A}\otimes\mathbb{I})^{*}_w(\mathbb{I}\otimes\hat{P_B})_w +r ((\hat{P_A}\otimes\hat{P_B})_w+ r(\hat{P_A}\otimes\mathbb{I})^{*}_w( \mathbb{I}\otimes\hat{P_B})_w) + (\hat{P_A}\otimes\mathbb{I})^{*}_w (\hat{P_A}\otimes\hat{P_B})_w \cos(2g)]\langle \hat{\sigma_{1}\otimes\mathbb{I}} \rangle \\
\nonumber
&-&\Im[( \mathbb{I}\otimes\hat{P_B})_w(\hat{P_A}\otimes\hat{P_B})^{*}_w + 2(\hat{P_A}\otimes\mathbb{I})_w + 2 r (\hat{P_A}\otimes\mathbb{I})_w( \mathbb{I}\otimes\hat{P_B})^{*}_w +r ((\hat{P_A}\otimes\hat{P_B})_w+r( \mathbb{I}\otimes\hat{P_B})_w(\hat{P_A}\otimes\hat{P_B}))^{*}\\
\nonumber
&+& ( \mathbb{I}\otimes\hat{P_B})^{*}_w (\hat{P_A}\otimes\hat{P_B})_w \cos(2g)]\langle \mathbb{I}\otimes\hat{\sigma_{2}} \rangle +[|\mathbb{I}-((\hat{P_A}\otimes\mathbb{I})_w+( \mathbb{I}\otimes\hat{P_B})_w)r + (\hat{P_A}\otimes\hat{P_B})_w r^{2}|^{2}+|{\hat{P_A}\otimes\mathbb{I}}_w|^{2}\\
\nonumber
&+& |({ \mathbb{I}\otimes\hat{P_B}})_w|^{2}+2 r^{2} |(\hat{P_A}\otimes\hat{P_B})_w|^{2}-2\Re[((\hat{P_A}\otimes\mathbb{I})_w+(\mathbb{I}\otimes\hat{P_B})_w)(\hat{P_A}\otimes\hat{P_B})_w]r]\langle \hat{\sigma_{1}}\otimes\hat{\sigma_{2}}\rangle \Big]W^{-1}_{5}.
\end{eqnarray}
\end{widetext}
where $r=1-\cos(g)$ and $W_5$ is normalization constant given in Eq.$(\ref{A2})$. Since, Eq.$(\ref{Mq})$ contains $ (\hat{P_A}\otimes\hat{P_B})_{w}$, we can extract joint weak value for all-order-coupling for qubit meter. We use  Eq.(\ref{Mq}) in the next section for analyzing the Hardy paradox when pointer state is taken to be a qubit. \\

\section{Application of rigorous treatment to the Hardy paradox}

The all-order-coupling treatment of the joint weak measurements of two observables by using continuous and discrete probe states can be used to re-examine in detail the well-known Hardy paradox. A typical Hardy paradox setup consists of two overlapping Mach-Zehnder interferometers(MZI), one for electron and other for positron. If the two MZI are kept separately then the positron is detected at $D_1$ and electron is at $D_4$. If the two MZIs are superposed, an overlapping region is created so that the presence of electron may disturb the interference effect of the interferometer $A$ and vice-versa. If the disturbance occurs then the positron may be detected at $D_2$ and electron may be detected at $D_3$.
\begin{figure}[h]
{\rotatebox{0}{\resizebox{8.0cm}{6.0cm}{\includegraphics{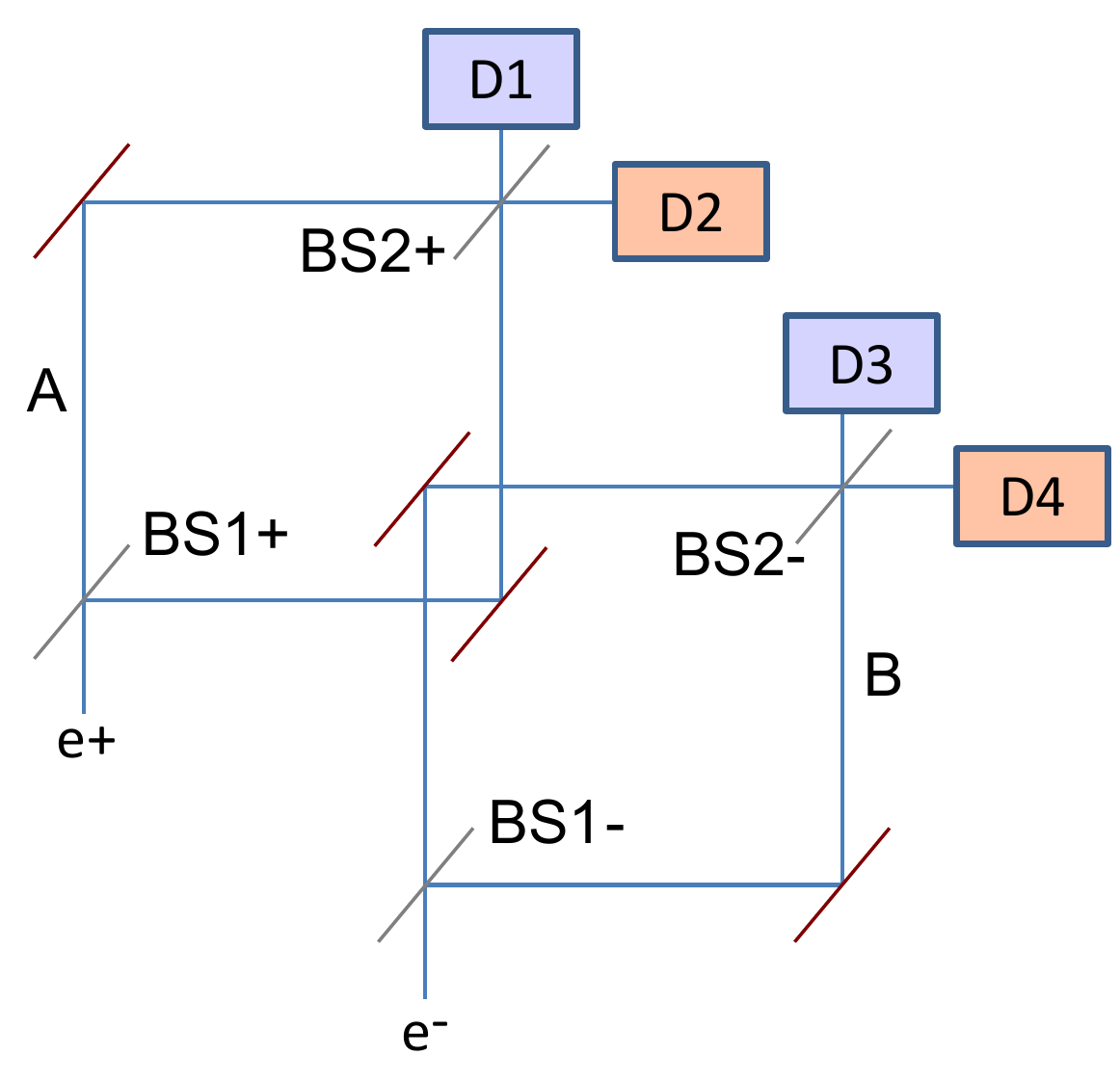}}}}
\caption{The Hardy Set-up}
\end{figure} 
 
The \textit{overlapping} and \textit{non-overlapping} paths are denoted by $|O\rangle$ and $|NO\rangle$ respectively. Due to the existence of this overlapping region there is a chance of annihilation, for that both electron and positron should be in overlapping arm, resulting neither detectors $D_2$ and $D_3$ click. If we try to understand the clicks at $D_{2}$ and $D_{3}$, it is found that to get click at $D_{3}$, positron must have to pass through overlapping arm but in order to avoid annihilation, electron should pass through non-overlapping arm. In the same way to get click at $D_{2}$ electron should have pass through overlapping arm but positron through non-overlapping arm to avoid annihilation. If simultaneous detection at $D_{2}$ and $D_{3}$ is obtained then both electron and positron have to travel through the overlapping arm. But in that case they should annihilate, hence the paradox. However, the whole argument is counterfactual. If actual measurement is performed the paradox disappear. But this counterfactual paradox can be explained if one performs non-local joint weak measurement. From the set-up, the normalized pre-selected state after the overlapping region turns out to be,
\begin{equation}
\label{pre}
|\phi_{i}\rangle = \frac{1}{\sqrt{3}}(|O_{A}\rangle|NO_{B}\rangle + |NO_{A}\rangle|O_{B}\rangle + |NO_{A}\rangle|NO_{B}\rangle)
\end{equation}
and in order to get joint detection at $D_2$ and $D_3$ the post-selected state is
\begin{equation}
\label{post}
|\phi_{f}\rangle = \frac{1}{2}(|O_{A}\rangle-|NO_{A}\rangle)(|O_{B}\rangle-|NO_{B}\rangle)
\end{equation}
 To inspect in which path electron and positron are going, one can invoke following projectors are given by,\\
\begin{eqnarray}
&&P_{O_{A}} = |O_{A}\rangle\langle O_{A}|; \ \  P_{NO_{A}} = |NO_{A}\rangle\langle NO_{A}|\\
\nonumber
&&P_{O_{B}} = |O_{B}\rangle\langle O_{B}|; \ \ P_{NO_{B}} = |NO_{B}\rangle\langle NO_{B}|
\end{eqnarray}
where $P_{O_{A}}$ denotes the path projector at the overlapping arm of the interferometer A and similarly for others. 
In order to find simultaneous presence of the electron and positron the joint projectors are defined as

\begin{eqnarray}
\nonumber
&&P_{O_{A}}\otimes P_{O_{B}}=|O_{A}\rangle\langle O_{A}|\otimes|O_{B}\rangle\langle O_{B}| \\  \nonumber
&&P_{O_{A}}\otimes P_{NO_{B}} = |O_{A}\rangle\langle O_{A}|\otimes|NO_{B}\rangle\langle NO_{B}|\\ \nonumber
&&P_{NO_{A}}\otimes P_{O_{B}} = |NO_{A}\rangle\langle NO_{A}|\otimes|O_{B}\rangle\langle O_{B}| \\ 
&&P_{NO_{A}}\otimes P_{O_{B}} = |NO_{A}\rangle\langle NO_{A}|\otimes|O_{B}\rangle\langle O_{B}|\\ \nonumber
\end{eqnarray}
Now, for the case of usual von Neumann measurement scenario, the expectation value of projector (say, $|A\rangle\langle A|$) provides the probability of finding the particle in that particular state $|A\rangle$. The pointer shift (say, $\delta x$) is dependent on coupling constant $(g)$ times the eigenvalue, i.e.,  $\delta x \in [0, g]$. Then the probability of finding the particles in the state $|A\rangle$ can be written as $Prob(|A\rangle)=\delta x/g$. If $\delta x =g$, the probability $Prob(|A\rangle)=1$. For the case of weak measurement aided with post-selection, the weak value of a projector can be called as weak probability which is actually a quasi-probability. Because, weak probability can even be negative. The interpretation of negative probability is then that the pointer is shifted to the opposite direction.  

In the Hardy setup, the weak values of projectors for the pre-selected state $|\phi_{i}\rangle$ given by Eq.$(\ref{pre})$ and post-selected state $|\phi_{f}\rangle$ given by Eq.$(\ref{post})$ are
 
\begin{eqnarray}
\label{single}
&&(P_{O_{A}})_{w}=(P_{O_{B}})_{w} = 1 \\
\nonumber
&&(P_{NO_{A}})_{w} =(P_{NO_{B}})_{w} = 0
\end{eqnarray}

Similarly for the weak values of the joint projectors given by

\begin{eqnarray}
\label{joint}
&&(P_{O_{A}}\otimes P_{O_{B}})_{w} = 0,  (P_{NO_{A}}\otimes P_{NO_{B}})_{w} = -1\\
\nonumber
&&(P_{O_{A}}\otimes P_{NO_{B}})_{w} =(P_{NO_{A}}\otimes P_{O_{B}})_{w} = 1 
\end{eqnarray}

From Eq.$(\ref{single})$ one concludes that the weak probability of finding the electron and positron independently in the overlapping arm is $1$.  This is required to obtain the detections of position and electron in $D_2$ and $D_3$ respectively. But in such a case they should annihilate. Interestingly, from Eq.$(\ref{joint})$ it is seen that joint weak probability $(P_{O_{A}}\otimes P_{O_{B}})_{w} = 0$, and hence condition to avoid annihilation is achieved. From Eq.(\ref{single}) we can also conclude that electron clicks at $D_2$ if and only if positron is in overlapping arm and to avoid annihilation, electron must be in non overlapping arm. This condition is achieved by noting that $(P_{O_{A}})_{w} = 1$, $(P_{NO_{B}})_{w} = 0$ and  $(P_{O_{A}}\otimes P_{NO_{B}})_{w}=1$. Similarly $(P_{NO_{A}})_{w} = 0$, $(P_{O_{B}})_{w} = 1$ and $(P_{NO_{A}}\otimes P_{O_{B}})_{w} = 1$ is explained as electron must be in overlapping arm otherwise positron will not click at $D_3$. We are getting two trajectories for the same particle at the same time, which is contradictory statement. However, we have $(P_{NO_{A}}\otimes P_{NO_{B}})_{w} = -1$, from which we can say that there is a minus one pair of electron and positron in non overlapping arm such that we will get only one pair of electron-positron at the end and that resolves the paradox. 

We now provide a detail analysis of the Hardy setup for all-order-coupling treatment to discuss how negative weak probability emerges in Hardy setup and to show how the mean pointer position changes with the coupling constant so that negative weak probability provide the shift in negative direction. In order to showing this, we have chosen a suitable pointer observable $\hat{Q}$ , so that, at the first order expansion of the $g$ (for joint measurement of two projectors, it is $g^2$) we obtain the weak joint probability is proportional to $\delta Q/g^2$. 

\begin{figure}[h]
{\rotatebox{0}{\resizebox{6.5cm}{4.5cm}{\includegraphics{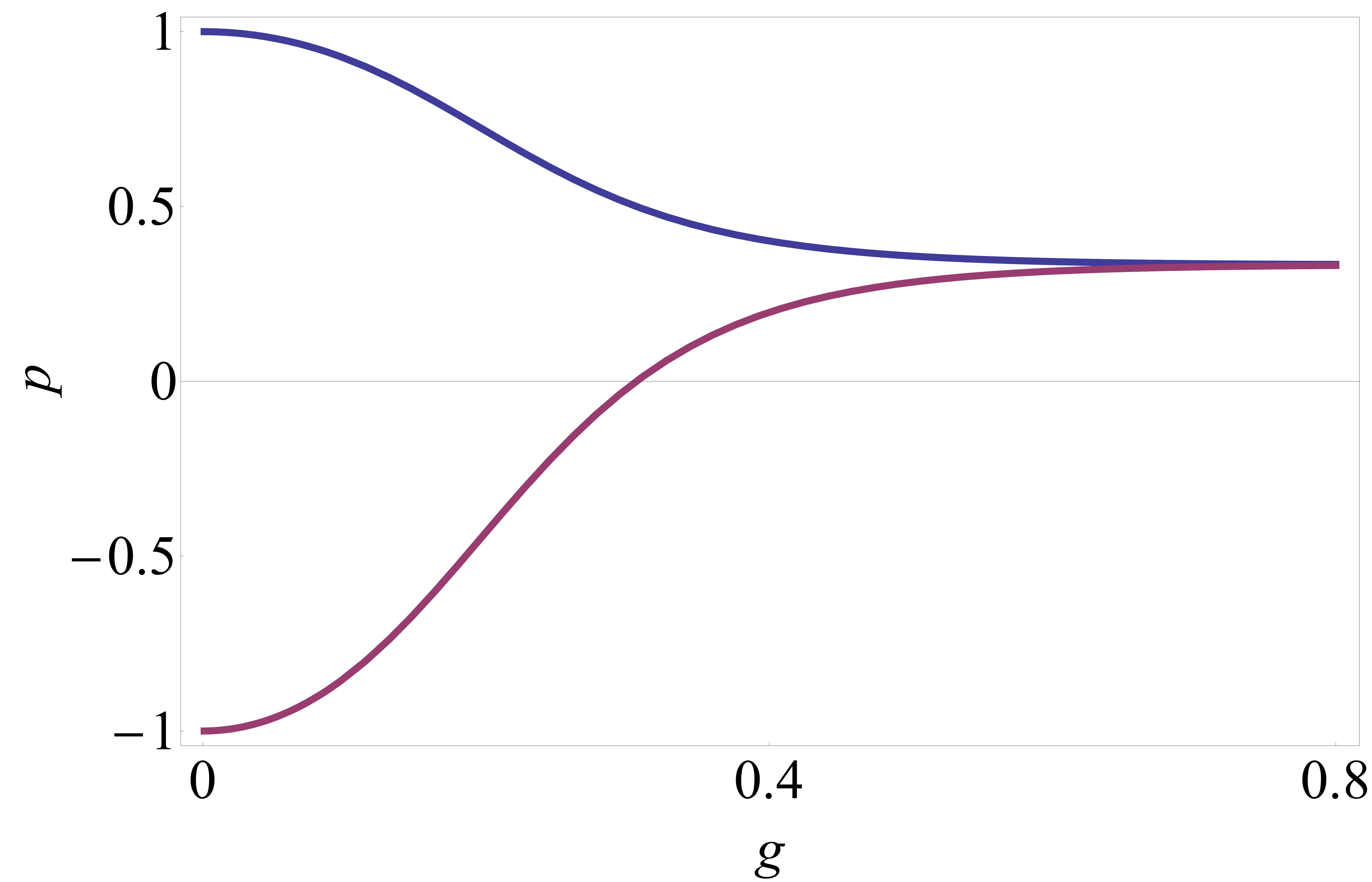}}}}
\caption{The probabilities given by Eq.$(\ref{eq6})$ and Eq.$(\ref{eq7})$ are plotted against g.}
\end{figure}
\subsection{Using continuous meter state}
For this, let us consider the mean pointer shift of a suitable joint observable, $\langle\delta Q\rangle = \langle\hat X \hat Y\rangle_{fi}-\sigma^{4}\langle\hat P_{x} \hat P_{y}\rangle_{fi}$. By using the expressions of $\langle\hat{X}\hat{Y}\rangle_{fi}$ and $\langle\hat P_{x}\hat P_{y}\rangle_{fi}$ that are already derived in Eq.$(\ref{xyp})$ and Eq.$(\ref{p1p2})$ respectively the quantity $\langle\delta Q\rangle$ can be calculated. The joint weak probability related to mean pointer shift of the joint observable for four different cases defined as $P_i=\frac{\langle\delta Q\rangle}{g^2}$ $(i=1,2,3,4)$ are the following; \\

$(i)$  $(P_{{O}_{A}})_{w} = (P_{{O}_{B}})_{w}=1$ and $(P_{{O}_{A}}\otimes P_{{O}_{B}})_{w} = 0$,
\begin{equation} \label{eq4}
P_{1} = 0 ,
\end{equation}
  $(ii)$ $(P_{{O}_{A}})_{w} = 1$, $ (P_{{NO}_{B}})_{w}=0$ and $(P_{{O}_{A}}\otimes P_{{NO}_{B}})_{w} = 1$, we have
\begin{equation} \label{eq5}
 P_{2}=\frac{ 1- e^{\frac{g^2}{4 \sigma ^2}}+ e^{\frac{g^2}{2 \sigma ^2}}}{ 2-4 e^{\frac{g^2}{4 \sigma ^2}}+3 e^{\frac{g^2}{2 \sigma ^2}}},
\end{equation}
 $(iii)$ $(P_{{NO}_{A}})_{w} = 0$,$ (P_{{O}_{B}})_{w}=1$ and $(P_{{NO}_{A}}\otimes P_{{O}_{B}})_{w} = 1$, the probability is given by
\begin{equation} \label{eq6}
P_{3}=\frac{1- e^{\frac{g^2}{4 \sigma ^2}}+ e^{\frac{g^2}{2 \sigma ^2}}}{ 2-4 e^{\frac{g^2}{4 \sigma ^2}}+3 e^{\frac{g^2}{2 \sigma ^2}}} ,
\end{equation}
and\\
 $(iv)$ $(P_{{NO}_{A}})_{w} = (P_{{NO}_{B}})_{w}=0$ and $(P_{{NO}_{A}}\otimes P_{{NO}_{B}})_{w} = -1$, the probability in this case is given by
\begin{equation} \label{eq7}
 P_{4}= \frac{ e^{\frac{g^2}{2 \sigma ^2}}-2 e^{\frac{g^2}{4 \sigma ^2}} }{ 2-4 e^{\frac{g^2}{4 \sigma ^2}}+3 e^{\frac{g^2}{2 \sigma ^2}}} .
\end{equation}

The probabilities given in Eq.$(\ref{eq6})$ and Eq.$(\ref{eq7})$ are plotted (Figure $2$) to show that how negative weak probability emerges at small coupling limit.

As the coupling strength of $g$ increases, weak probabilities gradually tending to become positive and increasing $g$ further the weak probabilities for case$(iii)$ and $(iv)$ both conditions overlap to each other. 
In weak coupling limit when we retain only upto the second order, Eqs.$(\ref{eq4}-\ref{eq7})$ reduce to,
\begin{equation} \label{eq8}
{(P_{1})}_w= 0, \\\ {(P_{2})}_w=1, \\\ {(P_{3})}_w=1 \\\ and \\\ {(P_{4})}_w= -1.
\end{equation} 
This clearly shows that for smaller value of $g$, how we can explain the Hardy paradox but as strength of coupling increase paradox disappears. 

\subsection{Using discrete meter state}
We now calculate the joint weak value by taking the qubit pointer in the context of Hardy set-up. For this, we consider the pointer state given as $|\xi_{i}\rangle = (1,0,0,0)^{T}$ and pointer observables as $(\hat\sigma_{x}\otimes\mathbb{I})$ and $(\mathbb{I}\otimes\hat\sigma_{x})$ . For such particular choice, the post-interaction pointer state is given as $|\xi_{f}\rangle=\langle\chi_{f}|\chi_{i}\rangle \eta_{2} |\xi_{i}\rangle$, where $\eta_{2} =1-(\mathbb{I}\otimes\hat{P_A})_{w} (1-\cos \left(g\right)+i (\hat\sigma_{x}\otimes\mathbb{I}) \sin \left(g\right))-(\hat{P_B}\otimes\mathbb{I})_{w}(1-\cos \left(g\right)+i (\mathbb{I}\otimes\hat\sigma_{x}) \sin \left(g\right))+(\hat{P_A}\otimes\hat{P_B})_{w}(1-\cos \left(g\right)+i (\hat\sigma_{x}\otimes\mathbb{I}) \sin \left(g\right))(1-\cos \left(g\right)+i (\mathbb{I}\otimes\hat\sigma_{x}) \sin \left(g\right))$.
 The mean value of a pointer observable, $\hat M = \frac{( \hat{\sigma_{x}}-\hat{\sigma_{y}})}{\sqrt{2}}\otimes\frac{(\hat{\sigma_{x}}+\hat{\sigma_{y}})}{\sqrt{2}}$ on the post-selected state $|\xi_{f}\rangle$ is calculated as,
\begin{widetext}
\begin{eqnarray}
\left\langle\frac{( \hat{\sigma_{x}}-\hat{\sigma_{y}})}{\sqrt{2}}\otimes\frac{(\hat{\sigma_{x}}+\hat{\sigma_{y}})}{\sqrt{2}}\right \rangle_{fi} &=&-2(\Im[2(\hat{P_A}\otimes\mathbb{I})_w( \mathbb{I}\otimes\hat{P_B})^{*}_w-(\hat{P_A}\otimes\mathbb{I})_w(\hat{P_A}\otimes\hat{P_B})^{*}_w-( \mathbb{I}\otimes\hat{P_B})_w(\hat{P_A}\otimes\hat{P_B})^{*}_w]\\
\nonumber
&+&\Re[(\hat{P_A}\otimes\hat{P_B})_w-(\hat{P_A}\otimes\mathbb{I})_w(\hat{P_A}\otimes\hat{P_B})^{*}_w-( \mathbb{I}\otimes\hat{P_B})_w(\hat{P_A}\otimes\hat{P_B})^{*}_w]\\
\nonumber
&+&(\Re[(\hat{P_A}\otimes\mathbb{I})_w(\hat{P_A}\otimes\hat{P_B})^{*}_w+( \mathbb{I}\otimes\hat{P_B})^{*}_w(\hat{P_A}\otimes\hat{P_B})^{*}_w]+\Im[(\hat{P_A}\otimes\mathbb{I})_w(\hat{P_A}\otimes\hat{P_B})^{*}_w\\
\nonumber
&+&(\hat{P_A}\otimes\mathbb{I})_w(\hat{P_A}\otimes\hat{P_B})^{*}_w]+2|(\hat{P_A}\otimes\hat{P_B})_w|^2)\cos(g)+|(\hat{P_A}\otimes\hat{P_B})_w|^2 \cos(2g)))\sin^2(g) W^{-1}_{6},
\end{eqnarray}
\end{widetext}

where $W_{6}$ is normalization constant given in Eq.$(\ref{A3})$.\\

Similar to the continuous meter state, we pointed out four different cases for the suitable choice of the joint observable defined by $P_{j}=\langle\frac{( \hat{\sigma_{x}}-\hat{\sigma_{y}})}{\sqrt{2}}\otimes\frac{(\hat{\sigma_{x}}+\hat{\sigma_{y}})}{\sqrt{2}} \rangle_{fi}/g^2$ where $j=1,2,3,4$, so that for \\

$(i)$  $(P_{{O}_{A}})_{w} = (P_{{O}_{B}})_{w}=1$ and $(P_{{O}_{A}}\otimes P_{{O}_{B}})_{w} = 0$ 
\begin{equation} \label{eq9}
P_{1} = 0 ,
\end{equation}
 $(ii)$  $(P_{{O}_{A}})_{w} = 1$  \  , \  $ (P_{{NO}_{B}})_{w}=0$ and $(P_{{O}_{A}}\otimes P_{{NO}_{B}})_{w} = 1$
\begin{equation} \label{eq10}
P_{2}=\frac{(2 \cos (g)-\cos (2 g)-3)}{ (8 \cos (g)-3 \cos (2 g)-7)}, 
\end{equation}
 $(iii)$  $(P_{{NO}_{A}})_{w} = 0$,$ (P_{{O}_{B}})_{w}=1$ and $(P_{{NO}_{A}}\otimes P_{{O}_{B}})_{w} = 1$  
\begin{equation} \label{eq11}
P_{3}=\frac{(2 \cos (g)-\cos (2 g)-3)}{ 8 \cos (g)-3 \cos (2 g)-7},
\end{equation}\\
and\\
 $(iv)$ $(P_{{NO}_{A}})_{w} = (P_{{NO}_{B}})_{w}=0$ and $(P_{{NO}_{A}}\otimes P_{{NO}_{B}})_{w} = -1$, we have  
\begin{equation} \label{eq12}
P_{4}=\frac{ (4 \cos (g)-\cos (2 g)-1)}{ 8 \cos (g)-3 \cos (2 g)-7}.
\end{equation}

If we take upto the first order of $g$,  from Eq.$(\ref{eq9}-\ref{eq12})$ reduce to the following

\begin{equation} 
\label{eq13}
{(P_{1})}_w= 0, \\\ {(P_{2})}_{w}=1, \\\ {(P_{3})}_{w}=1 \\\ and \\\ {(P_{4})}_{w}= -1.
\end{equation}
The above result is similar to the continuous case. Hence we showed that how negative probability emerges for both the Gaussian and the qubit pointers.\\

\section{Summary and Conclusions}
In this work we presented a rigorous analysis of joint weak measurement of two commutating observable for all-order-coupling by using continuous and discrete meter state. Specifically the type of observables we consider satisfy $A^2=\mathbb{I}$ and $A^2=A$. Note that, the joint weak value is usually calculated by restricting the coupling strength upto second order. We extend the study of joint weak measurement for all-order-coupling for showing that the joint weak value can be extracted for any order of coupling. The known result can be recovered in the second order of coupling strength. The rigorous treatment presented here enables us to obtain an interesting feature that the single meter displacement can provide the information of the joint weak value, if at least third order expansion of the coupling is invoked - a feature, which cannot be obtained, if only second order expansion of the coupling is taken as is done in earlier works. We also showed that the imaginary joint weak value can also be extracted by considering the statistics of single pointer instead of joint pointer observables. As an application, we re-examined the well known Hardy paradox and provide an all-order-coupling treatment for the same by using discrete and continuous pointer. Such a treatment allow us to see how the negative weak probability emerges at the weak coupling limit. Finally, since all-order-coupling treatment lifted the constraints on the strength of the coupling, the results presented here can be helpful for experimentally testing the joint weak value.
\section*{Acknowledgments}
AKP acknowledges the support from Ramanujan Fellowship research grant (SB/S2/RJN-083/2014). 

\begin{widetext}
\appendix
\section{}

The explicit expressions of the quantities $W_{3}$, $W_{5}$ and $W_{6}$ that are used in the main text are respectively given by

\begin{eqnarray}
\label{A1}
W_{3}&=&   e^{\frac{g^2}{4 \sigma ^2}} +2((\hat{P_A}\otimes\mathbb{I})_{w}-(\hat{P_A}\otimes\mathbb{I})_{w}(\mathbb{I}\otimes\hat{P_B})^*_{w}+|(\hat{P_A}\otimes\hat{P_B})_{w}|^2)e^{\frac{g^2}{8 \sigma ^2}}\\
\nonumber
&+&2(|(\mathbb{I}\otimes\hat{P_B})_{w} |^2+3(\hat{P_A}\otimes\hat{P_B})_{w}+2|(\hat{P_A}\otimes\hat{P_B})_{w}|^2-2(\mathbb{I}\otimes\hat{P_B})_{w} (\hat{P_A}\otimes\hat{P_B})^*_{w})(1 - e^{\frac{g^2 }{8\sigma^2}})^2+2(\hat{P_B})_{w}(e^{\frac{g^2 }{8\sigma^2}} - e^{\frac{g^2 }{4 \sigma^2}})
\end{eqnarray}

\begin{eqnarray}
\label{A2}
W_{5} &=& \Big[|1-((\hat{P_A}\otimes\mathbb{I})_w+(\mathbb{I}\otimes\hat{P_B})_w)r + (\hat{P_A}\otimes\hat{P_B})_w r^{2}|^{2}+|({\hat{P_A}\otimes\mathbb{I}})_w|^{2}+|(\mathbb{I}\otimes{\hat{P_B}})_w|^{2}+2 r^{2} |(\hat{P_A}\otimes\hat{P_B})_w|^{2}\\
\nonumber
&-&2\Re[((\hat{P_A\otimes\mathbb{I}})_w+(\mathbb{I}\otimes\hat{P_B})_w)(\hat{P_A}\otimes\hat{P_B})_w]r]-\Im[(\hat{P_B})_w(\hat{P_A}\otimes\hat{P_B})^{*}_w +2(\hat{P_A}\otimes\mathbb{I})_w + 2 r (\hat{P_A}\otimes\mathbb{I})_w(\mathbb{I}\otimes\hat{P_B})^{*}_w  \\
\nonumber
&+& r ((\hat{P_A}\otimes\hat{P_B})_w+r(\mathbb{I}\otimes\hat{P_B})_w(\hat{P_A}\otimes\hat{P_B})^{*}_w)+(\mathbb{I}\otimes\hat{P_B})^{*}_w (\hat{P_A}\otimes\hat{P_B})_w \cos(2g)]\langle \hat{\sigma_{1}\otimes\mathbb{I}} \rangle\\
\nonumber
&-& \Im[(\hat{P_A}\otimes\mathbb{I})^{*}_w(\hat{P_A}\otimes\hat{P_B})_w+2(\mathbb{I}\otimes\hat{P_B})_w + 2 r (\hat{P_A}\otimes\mathbb{I})^{*}_w(\mathbb{I}\otimes\hat{P_B})_w  +r ((\hat{P_A}\otimes\hat{P_B})_w +r(\hat{P_A}\otimes\mathbb{I})^{*}_w(\mathbb{I}\otimes\hat{P_B})_w)\\
\nonumber
&+& (\hat{P_A}\otimes\mathbb{I})^{*}_w (\hat{P_A}\otimes\hat{P_B})_w \cos(2g)]\langle \mathbb{I}\otimes\hat{\sigma_{2}} \rangle+ 2\Re[(\hat{P_A}\otimes\mathbb{I})_w(\mathbb{I}\otimes\hat{P_B})^{*}_w + (\hat{P_A}\otimes\hat{P_B})_w\Big]\sin ^2(g)\langle \hat{\sigma_{1}}\otimes\hat{\sigma_{2}}\rangle
\end{eqnarray}

\begin{eqnarray}
\label{A3}
W_{6}&=&|(1-(\hat{P_A}\otimes\mathbb{I})_{w}-(\mathbb{I}\otimes\hat{P_B})_{w}+(\hat{P_A}\otimes\hat{P_B})_{w}+\cos(g((\hat{P_A}\otimes\mathbb{I})_{w}+(\mathbb{I}\otimes\hat{P_B})_{w}-2(\hat{P_A}\otimes\hat{P_B})_{w}\\
\nonumber
&+&(\hat{P_A}\otimes\hat{P_B})_{w}\cos(g))))|^2+|((\hat{P_A}\otimes\mathbb{I})_{w}-(\hat{P_A}\otimes\hat{P_B})_{w}+(\hat{P_A}\otimes\hat{P_B})_{w}\cos(g))|^2 \sin^2(g)\\
\nonumber
&+&|((\mathbb{I}\otimes\hat{P_B})_{w}-(\hat{P_A}\otimes\hat{P_B})_{w}+(\hat{P_A}\otimes\hat{P_B})_{w}\cos(g))|^2 \sin^2(g)+|(\hat{P_A}\otimes\hat{P_B})_{w}|^2\sin^4(g)
\end{eqnarray}
\end{widetext}
\end{document}